%% file: main.tex
\documentclass{Interspeech}

\interspeechcameraready 

\title{ToxicTone: A Mandarin Audio Dataset Annotated for \\ Toxicity and Toxic Utterance Tonality}

\author[affiliation={1}]{Yu-Xiang}{Luo$^*$}
\author[affiliation={1}]{Yi-Cheng}{Lin$^*$}
\author[affiliation={1}]{Ming-To}{Chuang$^*$}
\author[affiliation={1}]{Jia-Hung}{Chen}
\author[affiliation={1}]{I-Ning}{Tsai}
\author[affiliation={1}]{Pei Xing}{Kiew}
\author[affiliation={1}]{Yueh-Hsuan}{Huang}
\author[affiliation={1}]{Chien-Feng}{Liu}
\author[affiliation={1}]{Yu-Chen}{Chen}
\author[affiliation={1}]{Bo-Han}{Feng}
\author[affiliation={1}]{Wenze}{Ren}
\author[affiliation={1}]{Hung-yi}{Lee}

\affiliation{}{National Taiwan University}{Taiwan}
\email{\{b10902037, f12942075, r13942091, hungyilee\}@ntu.edu.tw}

\usepackage{comment}
\usepackage{hyperref}
\usepackage{url}
\usepackage{pgfplots}
\usepackage{subcaption}

\begin{document}

\maketitle

\begingroup
  \renewcommand\thefootnote{*}
  \footnotetext{Equal contribution}
\endgroup

\begin{abstract}
   Despite extensive research on toxic speech detection in text, a critical gap remains in handling spoken Mandarin audio. The lack of annotated datasets that capture the unique prosodic cues and culturally specific expressions in Mandarin leaves spoken toxicity underexplored. To address this, we introduce ToxicTone—the largest public dataset of its kind—featuring detailed annotations that distinguish both forms of toxicity (e.g., profanity, bullying) and sources of toxicity (e.g., anger, sarcasm, dismissiveness). Our data, sourced from diverse real-world audio and organized into 13 topical categories, mirrors authentic communication scenarios. We also propose a multimodal detection framework that integrates acoustic, linguistic, and emotional features using state-of-the-art speech and emotion encoders. Extensive experiments show our approach outperforms text-only and baseline models, underscoring the essential role of speech-specific cues in revealing hidden toxic expressions.
\end{abstract}

{\color{red} Warning: This paper may contain uncomfortable content.}
\vspace{-3pt}

\section{Introduction}
Toxic speech in online media is a serious global problem that creates hostile environments, discourages participation, and silences vulnerable voices. For individuals, exposure to toxic speech can cause psychological problems like stress, anxiety, and depression \cite{kowalski2014bullying}. This concern is more serious on social platforms, where users—especially those from marginalized groups—are often targeted by harmful language, resulting in exclusion and emotional harm \cite{bliuc2018online}. Over time, repeated exposure can cause long-term mental health issues like trauma and reduced self-esteem.

Although previous research has focused on toxicity detection in English and other well-supported languages, progress for Mandarin Chinese remains slow, mainly because large labeled datasets are scarce \cite{deng-etal-2022-cold}. In addition, most work focuses on text input and does not address the unique challenges of spoken communication (for example, voice chats and audio streams). Furthermore, Mandarin includes unique colloquialisms and code-switching influenced by local languages \cite{Chiu2012Code}, meaning that detection models must account for culture-specific toxic expressions \cite{rtplx}. Existing Chinese datasets also often lack detailed labels (for instance, separating toxic speech from general profanity or marking hidden insults \cite{lu-etal-2023-facilitating}), limiting the creation of culturally aware and strong toxicity detection solutions.

In this paper, we tackle these problems by building a large Mandarin Chinese toxic speech dataset, ToxicTone, which we believe is the biggest public resource of its kind. Unlike past datasets, ours includes detailed labels on the source of toxicity, capturing not only explicit or hateful words but also prosodic features (intonation, emphasis, rhythm) that show harmful intent yet cannot be seen from text alone. By showing how aggression emerges through vocal cues, ToxicTone uncovers toxic content that may be hidden behind seemingly polite words.

We also show that combining multiple model architectures trained on our dataset leads to better performance in toxic speech detection than text-only approaches. In particular, combining speech, emotion, and semantic encoders achieves the best results. This finding highlights the need to include prosodic and emotional information in speech, and supports the idea that we need special, speech-focused methods.

Our main contributions are as follows:
\begin{itemize}
    \item We release the largest public Mandarin Chinese toxic speech detection dataset, with detailed labels on both the form and the source of toxicity\footnote{\tiny https://github.com/YuXiangLo/ToxicTone}.
    \item We demonstrate that combining acoustic, emotional, and semantic features significantly boosts toxicity detection accuracy, showing that speech data is essential for dealing with toxic content.
\end{itemize}

By filling these data and modeling gaps, we aim to support safer online spaces and advance spoken-language research on toxicity detection, both for Mandarin Chinese and for broader uses in speech analytics.

\vspace{-3pt}

\section{Related work}
Previous works on Chinese toxic speech detection focus mainly on text. 
COLA \cite{tang-shen-2020-categorizing} represents the first Chinese offensive language classification dataset, comprising 18.7k comments sourced from YouTube and Weibo. The dataset categorizes texts into four classes: neutral, insulting, antisocial, and illegal. TOCP \cite{yang-lin-2020-tocp}, which focuses on Chinese profanity, contains 16k toxic comments collected from the PTT Bulletin Board and Twitch livestream chatrooms. Similarly, COLD \cite{deng-etal-2022-cold} serves as the first Chinese offensive language detection benchmark, consisting of 37k offensive language samples and anti-bias content related to race, gender, and region. Lastly, ToxiCN \cite{lu-etal-2023-facilitating} provides 12k hierarchical annotation for texts from  Zhihu or Tieba, including toxic type, targeted group, and expression.

These datasets are limited to text-based content, which does not capture the complexity of spoken language. Unlike spoken utterances, written text lacks prosodic features such as intonation, pitch, and stress, which can convey subtle expressions of toxicity or sarcasm. Furthermore, text datasets from social platforms cannot reflect real-world interactions, especially spontaneous language, emotion, and interruption.

The currently available datasets for detecting toxic speech include Detoxy \cite{ghosh22b_interspeech}, Mutox \cite{costa-jussa-etal-2024-mutox}, and ADIMA \cite{9746718}. Detoxy labels subsets of pre-existing speech datasets, such as CMU-MOSI \cite{zadeh2016mosi}, Common Voice \cite{ardila2020common}, and Switchboard \cite{godfrey1992switchboard}, by determining whether the samples are toxic or non-toxic. Notably, all the samples in Detoxy are in English, which limits the dataset's ability to generalize to other languages. On the other hand, 
Mutox annotates multilingual segments derived from SeamlessAlign \cite{barrault2023seamless} and Common Voice. However, the dataset includes only 2,000 samples per language, and the audio primarily originates from podcast recordings. This focus limits its representation of real-world conversational scenarios, such as phone calls, live-streamed gameplay, or drama. The ADIMA dataset, focused on abuse detection in 10 Indic languages, provides a diverse multilingual approach but is limited to detecting profanity.
\vspace{-5pt}

\section{Dataset collection}
\vspace{-3pt}
\subsection{Definition of toxicity}
\vspace{-3pt}
We define toxicity via two aspects: the \textbf{form} of toxicity and the \textbf{source} of toxicity.

\textbf{Forms of toxicity:} These describe the specific manifestations of harmful or offensive language. They include:
\begin{itemize}
    \item Profanities (Prof.): Offensive words that demonstrate disrespect or negativity. For example, \textit{fuck}, \textit{bastard}, \textit{sissy}, \textit{idiot}.
    \item Hate speech (Hate): Language that explicitly or implicitly expresses hostility, discrimination, or hatred toward groups based on their inherent or perceived characteristics. For example, \textit{bitch}, \textit{XX party dog}.
    \item Pornographic language (Porn.\ Lang.): Language that refers to sexual acts or body parts in a vulgar way, primarily intended to arouse sexual interest or evoke shock. For example, \textit{cum}, \textit{tits}, \textit{boobs}, \textit{cock}, \textit{pussy}.
    \item Bullying speech (Bully): Threatening, offensive, or aggressive speech that does not target a specific group. For example, \textit{shut up}, \textit{you'll die}, \textit{you suck}.
    \item Sarcasm (Sarc.): Utterances that convey a meaning opposite or significantly different from the literal words used, often degrading or mocking the target. For example, \textit{You are so smart, you are a genius}.
    \item Other toxic speech (Oth.\ Tox.): Language that does not fit the above categories, but still makes the listener feel disgusted or uncomfortable.
\end{itemize}

\textbf{Sources of toxicity:} These refer to the origin of the harmful intent or tone in communication. They include:
\begin{itemize}
    \item Specific Words (Spec.): Use of explicitly aggressive or offensive words that carry direct insults or defamatory meanings.
    \item Angry or Violent Tone (Ang./Viol.): Speech that directly expresses anger with emotional or provocative content, potentially implying violent actions.
    \item Dismissive or Impatient Tone (Dism./Imp.): A tone marked by derogatory adjectives or dismissiveness, often appearing indifferent or unfriendly.
    \item Sarcastic or Satirical Tone (Sarc./Satir.): A condescending tone used to mock or ridicule the target, often through double entendre or implied meanings.
    \item Explicit/Implicit Threatening Tone (Threat.): Speech that directly or indirectly intimidates the target, causing mental or emotional fear.
\end{itemize}

\subsection{Preprocessing}
We used web-crawled speech data as the basis of our research. After downloading the audio recordings, we employed a speaker diarization model\footnote{\tiny https://huggingface.co/pyannote/speaker-diarization-3.1} to differentiate and segment speech from multiple speakers, following \cite{yang2024buildingtaiwanesemandarinspoken}. Subsequently, the segmented audio was transcribed into text using the K2D model \cite{tseng2024leave} and splitting the results into 2–10 second clips. 

Given the enormous number of generated segments, a preliminary filtering step was required. To this end, we applied a text-based toxicity classifier from Alibaba-pai\footnote{\tiny https://huggingface.co/alibaba-pai/pai-bert-base-zh-llm-risk-detection}—based on Chinese BERT-base \cite{devlin-etal-2019-bert} —to the transcriptions. The classifier assigns a toxicity score between 0 (no toxicity) and 1 (very toxic), and only segments scoring above 0.75 were retained for subsequent human annotation and analysis, because this value effectively balances the need to filter out non-toxic content while capturing segments that are likely to contain significant toxic elements.  This filtering reduced the total number of clips from 770k to 52k.

In addition, because our preliminary filtering seldom detected speech with explicit sexual content, we employed a rule-based word list\footnote{\tiny https://github.com/facebookresearch/flores/blob/main/toxicity} to extract an additional 600 samples that potentially exhibit pornographic toxicity. 

\subsection{Human Annotation}
The dataset was annotated by a team of 11 native Chinese annotators. The annotation task took approximately 1.5 months to complete, with annotators collectively spending around 900 hours. All annotators are informed that they might encounter uncomfortable audio content, and they can quit the task at any time. For each sample, annotators could select one or more forms of toxicity and their corresponding sources, or indicate that the sample was non‑toxic. Annotators would also filter out the audio not in Mandarin. Initially, each sample was annotated by four annotators. Samples receiving an equal number of ``toxic" and ``non-toxic" annotations were subsequently reviewed and annotated by an additional annotator to resolve discrepancies. 

The final dataset was separated into train, development, and test sets. The statistics of our dataset are depicted in Table~\ref{tab:dataset_summary} and Figure ~\ref{fig:combined_toxicity}. Compared to other datasets in Table~\ref{tab:dataset_comparison}, our dataset is the largest publicly available toxic speech detection dataset.

The labels are imbalanced in our dataset both in the forms and sources of toxicity. For example, there are nearly 5,800 bullying speech samples, while pornographic language has fewer than 400 samples. Similarly, samples labeled with specific words appear almost 8,000 times, while those with a threatening tone occur only about 560 times. This imbalance reflects the real-world distribution of toxic speech. Research shows that some forms of toxic language—such as casual profanity and bullying—are more common in everyday online interactions because they are often tolerated or even normalized in many communities \cite{cheng2015antisocial, lenhart2016online}. The samples of the ToxicTone dataset are in the supplementary material.

\input{Tables/statistic}
\input{Tables/comparison}
\input{plots/toxic_statistic}

\subsection{Category distribution}
Motivated by the observation that the topical focus of a speech segment can influence both the prevalence and expression of toxic language, we divide our dataset into 13 topical categories. These categories—Society \& News, Technology \& Science, Education, Gaming, Entertainment, Culture \& Arts, Psychology \& Lifestyle, Movie \& Book Reviews, Food, Health \& Fitness, Parenting \& Family, Beauty \& Fashion, and Business—are designed in line with the categorization systems used by Apple Podcasts\footnote{\tiny https://podcasters.apple.com/support/1691-apple-podcasts-categories} and Spotify Podcasts\footnote{\tiny https://open.spotify.com/genre/0JQ5DArNBzkmxXHCqFLx2U}. To assign each audio sample to one of these categories, we use GPT-4o mini \cite{openai2024gpt4ocard} to classify the samples in batches of 20, based on transcripts from the first 30 minutes of the spoken content before segmenting to 2-10 seconds.

Figure~\ref{fig:clip_counts} shows the number of clips in each category, highlighting the variety in our dataset. The largest groups are Society \& News and Entertainment, with 14,247 and 13,684 clips respectively, and they also contain a high number of toxic clips. Gaming also has strong representation with 8,326 total clips, 4,133 of which are toxic. On the other hand, smaller categories such as Beauty \& Fashion (589 clips with 135 toxic clips) and Movie \& Book Reviews (249 clips with 56 toxic clips) are well represented too. This varied distribution, covering both high-volume mainstream topics and more niche areas, highlights the diversity of our dataset and its value for studying toxic speech.


\input{plots/category}

\vspace{-5pt}
\section{Experiments}
\subsection{Experiment Type}
We evaluate two classification tasks. The first, toxicity detection, determines whether a given speech segment contains toxic content. This task is a binary classification problem, where the model outputs a score between 0 and 1, with a threshold-based decision to classify an utterance as toxic or non-toxic. 


The second task, toxicity source classification, identifies the origin of toxicity in speech. We classify toxic utterances into source categories, including aggressive wording, sarcastic tone, or threatening intent. Since toxicity in speech is often implicit, this task helps distinguish between overtly toxic expressions and subtler toxic cues embedded in speech patterns.
\vspace{-8pt}
\input{plots/toxicity_source_classification}
\subsection{Experiment Setup}
To evaluate toxicity detection in spoken Mandarin, we compare our approach against three baseline systems: MuTox, ETOX\footnote{\tiny https://github.com/facebookresearch/seamless\_communication/tree/main} \cite{etox}, and COLD\textsubscript{ETECTOR}\footnote{\tiny https://huggingface.co/thu-coai/roberta-base-cold} \cite{deng-etal-2022-cold}. MuTox is a multilingual speech-based toxicity classifier that employs a pre-trained speech encoder (SONAR) \cite{sonar} with a three-layer feedforward network for binary classification. It represents a state-of-the-art approach to speech toxicity detection. In contrast, ETOX is a lexicon-based toxicity detection system operating on text. It relies on predefined wordlists covering multiple languages to flag toxic content, making it highly interpretable. However, since it operates on explicit lexical cues, it struggles with context-dependent or implicit toxicity, which is common in spoken communication. Additionally, COLD\textsubscript{ETECTOR} is built upon bert-base-chinese and fine-tuned on a large-scale Chinese offensive language dataset, enabling it to capture both explicit and subtle offensive cues.

We investigate three pre-trained models to encode audio into features. SONAR (\textbf{S}) \cite{sonar} is a multilingual sentence embedding model supporting both text and audio inputs; we experiment with both its text encoder (applied to ASR-transcribed speech, $\mathbf{S_T}$)\footnote{\tiny https://dl.fbaipublicfiles.com/SONAR/sonar\_text\_encoder.pt} and speech encoder (applied to raw audio, $\mathbf{S_A}$)\footnote{\tiny https://dl.fbaipublicfiles.com/SONAR/spenc.v5ap.cmn.pt}. Additionally, we employ XLS-R 1B (\textbf{X})\footnote{\tiny https://dl.fbaipublicfiles.com/fairseq/wav2vec/xlsr2\_960m\_1000k.pt} \cite{xlsr}, a multilingual self-supervised speech model that captures rich prosodic and acoustic features; for XLS-R, we compute a layerwise weighted sum of its features to obtain a robust embedding \cite{yang21c_interspeech}. To incorporate emotional cues, we use Emotion2Vec+ Large (\textbf{E})\footnote{\tiny https://huggingface.co/emotion2vec/emotion2vec\_plus\_large} \cite{emotion2vec}, a speech emotion representation model pre-trained on large-scale emotion datasets, and we extract its last-layer embedding. These models provide complementary information, capturing semantic, acoustic, and emotional aspects of toxicity in spoken Mandarin. In addition, we explore ensemble models by concatenating the individual features along the feature dimension, allowing us to leverage the strengths of each model.

For downstream classification, all tasks utilize a three-layer linear classifier trained with binary cross-entropy loss. The toxicity detection model predicts a binary toxicity score. For toxicity source classification, we train separate classifiers per category using a one-vs-all approach, treating each class independently. This setup allows us to analyze the different manifestations of toxicity in a structured and interpretable manner.

\subsection{Toxicity detection}
\vspace{-3mm}
\input{Tables/toxicity_classification}
Table~\ref{tab:embedding_results} summarizes the classification performance across different models and feature combinations. Our model trained on our dataset outperforms all baselines.
Baseline performance shows that Mutox scores low, while ETOX attains higher performance, with COLD\textsubscript{ETECTOR} slightly lagging behind. Individually, the emotion-based embedding (\(E\)) scores an F1 of 47.96\%, suggesting that emotional signals alone are insufficient, whereas XLS-R (\(X\)) reaches an F1 of 60.89\% with balanced metrics, which is the best among all single encoder settings. The SONAR encoders are competitive, with \(S_T\) achieving an F1 of 57.49\% and a recall of 59.11\%. Notably, combining embeddings improves performance: \(S_T+E\) yields an F1 of 62.47\% and a recall of 70.09\%, and the multimodal \(X+S_T+E\) configuration attains the best results, underscoring the value of fusing acoustic, linguistic, and emotional features for effective toxic speech detection.

\subsection{Toxicity source detection}
The performance of toxicity source detection is depicted in Figure~\ref{fig:toxic_source_performance}. For the ``Specific Words'' category, all models exhibit consistently high F1 scores (approximately 0.73–0.75) and comparable accuracy, indicating that explicit lexical cues are reliably detected even with single model input. For categories like "Angry/Violent" and "Sarcastic/Satirical," the ensemble configuration ($S_T+E+X$) excels with the highest F1 scores and improved accuracy. In contrast, threat detection, despite uniformly high accuracy, yields a lower F1 score with the best performance achieved by the semantic encoder ($S_T$). This reduced performance can be attributed to the imbalanced distribution of threat-labeled samples, which are considerably fewer.

\vspace{-6pt}
\section{Limitation and future work}
\vspace{-2pt}
While our dataset and models establish a strong foundation for Chinese spoken toxic speech detection, several areas offer opportunities for further refinement. The dataset reflects real-world toxicity distributions, including natural class imbalances, such as the higher prevalence of toxic speech in gaming content. While this alignment with authentic usage patterns enhances model relevance, future work can expand coverage of underrepresented categories to improve generalization.


Like most toxic speech detection datasets, our experiments reflect majority opinion, which may not fully capture individual perceptions influenced by personal experience, culture, and context \cite{lin24i_interspeech, lin24b_interspeech, garg2023handling, 10832259}. 
To enhance transparency and enable further research on annotation biases, we will release the annotator-level annotations. 
Future work can leverage these detailed annotations to explore alternative aggregation strategies and develop adaptive models that better account for cultural and individual differences in toxicity perception.
\vspace{-6pt}
\section{Conclusion}
\vspace{-2pt}
This work introduces the first large-scale Mandarin Chinese toxic speech dataset, addressing a critical gap in spoken toxic speech detection. Unlike prior text-based datasets, our dataset incorporates prosodic cues and detailed toxicity labels, enabling a more nuanced understanding of harmful speech. Our experiments demonstrate that multimodal approaches—leveraging acoustic, emotional, and linguistic features—significantly outperform text-only models, underscoring the importance of speech-specific cues in detecting toxicity. Additionally, our analysis reveals domain-specific trends in toxicity, with gaming content exhibiting the highest prevalence. These findings highlight the necessity of dedicated speech-based detection models to capture the complexities of spoken toxicity. By establishing a strong benchmark for Chinese spoken toxic speech detection, our work lays the foundation for future advancements in multimodal toxicity detection, dataset expansion across Chinese dialects, and personalized toxicity perception models to further enhance content moderation in online speech interactions.

\bibliographystyle{IEEEtran}
\bibliography{mybib}

\end{document}

%% file: Tables/statistic.tex
\begin{table}[h!]
\centering
\caption{Statistic of ToxicTone. TL refers to the total length of the dataset, in format hh:mm:ss.}
\vspace{-4pt}
\begin{tabular}{@{}lcccc@{}}
\toprule
\textbf{Split} & \textbf{\# Utt.} & \textbf{\# Toxic} & \textbf{\# Non-Toxic} & \textbf{TL} \\ \midrule
Train          & 41,649 & 13,401 & 28,248 & 74:29:52 \\
Dev            & 5,206  & 1,654  & 3,552  & 9:16:49  \\
Test           & 5,207  & 1,672  & 3,535  & 9:22:09  \\ \midrule
\textbf{Total} & 52,062 & 16,727 & 35,335 & 93:08:50 \\ 
\bottomrule
\end{tabular}
\label{tab:dataset_summary}
\end{table}

%% file: Tables/comparison.tex
\begin{table}[h!]
\centering

\caption{Comparison with MuTox (English+Spanish), DeToxy-B, and ADIMA. TL refers to the total dataset length.}
\vspace{-6pt}
\begin{tabular}{@{}lcccc@{}}
\toprule
\textbf{Dataset} & \textbf{\# Utt.} & \textbf{\# Toxic} & \textbf{\# Non-Toxic} & \textbf{TL} \\ 
\midrule
DeToxy-B                  & 20,217 & 5,307  & 14,910 & 24:39:59 \\
MuTox                     & 40,000 & 7,143  & 31,919 & 43:12:00 \\
ADIMA & 11,775 & 5,108 & 6,667 & 65:00:00 \\
\textbf{ToxicTone}     & \textbf{52,062} & \textbf{16,727} & \textbf{35,335} & \textbf{93:08:50} \\
\bottomrule
\end{tabular}
\label{tab:dataset_comparison}
\end{table}

%% file: plots/toxic_statistic.tex
\begin{figure}[!ht]
\centering

\begin{subfigure}[t]{0.52\columnwidth}
    \centering
    \begin{tikzpicture}
        \begin{axis}[
            ybar,
            trim axis left,
            trim axis right,
            bar width=0.3cm,
            ylabel near ticks,
            width=1.05\columnwidth,
            height=4.2cm,
            ylabel={Clip Count},
            symbolic x coords={Prof., Hate, Porn. Lang., Bully, Sarc., Oth. Tox.},
            ylabel style={font=\scriptsize,yshift=-0.15cm},
            xtick=data,
            ymin=0,
            ymax=8500, 
            enlarge x limits=0.15,
            tick label style={font=\scriptsize},
            xticklabel style={rotate=45, anchor=east},
            xtick pos=lower,
        ]
            \addplot coordinates {
              (Prof., 3712)
              (Hate, 1074)
              (Porn. Lang., 350)
              (Bully, 5803)
              (Sarc., 1479)
              (Oth. Tox., 200)
            };
        \end{axis}
    \end{tikzpicture}
    \caption{Form of Toxicity}
    \label{fig:form_toxicity}
\end{subfigure}%
\hspace{-25pt}
\raisebox{0.9pt}{
\begin{subfigure}[t]{0.5\columnwidth}
    \centering
    \begin{tikzpicture}
        \begin{axis}[
            ybar,
            trim axis left,
            trim axis right,
            bar width=0.3cm,
            width=1.1\columnwidth,
            height=4.2cm,
            symbolic x coords={Spec. , Ang./Viol., Dism./Imp., Sarc./Satir., Threat.},
            xtick=data,
            yticklabels = {},
            ytick       = {0,2000,4000,6000,8000},
            ymin=0,
            ymax=8500, 
            enlarge x limits=0.2,
            tick label style={font=\scriptsize},
            xticklabel style={rotate=45, anchor=east},
            xtick pos=lower,
        ]
            \addplot coordinates {
              (Spec. , 7986)
              (Ang./Viol., 2080)
              (Dism./Imp., 3575)
              (Sarc./Satir., 2016)
              (Threat., 561)
            };
        \end{axis}
    \end{tikzpicture}
    \caption{Source of Toxicity}
    \label{fig:source_toxicity}
\end{subfigure}
}
\vspace{-4pt}
\caption{
    Comparison of clip counts by Form and Source of Toxicity.
}
\vspace{-10pt}
\label{fig:combined_toxicity}
\end{figure}
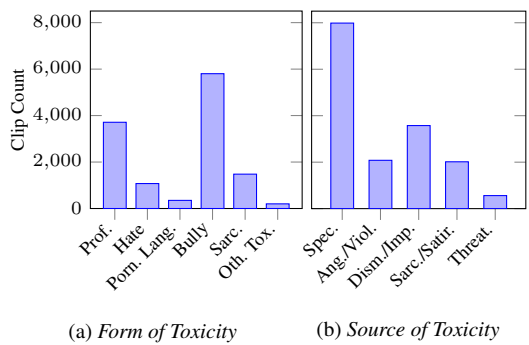

%% file: plots/category.tex
\begin{figure}[!ht]
    \centering
    \begin{tikzpicture}
        \begin{axis}[
            ybar,
            width=\columnwidth,
            height=4.8cm,
            bar width=0.3cm,
            ymin=0,
            ymax=15000, 
            ylabel={Clip Count},
            symbolic x coords={
                SocNews,
                Entr,
                Gaming,
                PsychLife,
                HealthFit,
                Educ,
                Food,
                CultArts,
                Bus,
                TechSci,
                ParentFam,
                BeautFash,
                MovBook
            },
            xtick=data,
            xticklabel style={rotate=45, anchor=east, font=\small},
            tick label style={font=\small},
            legend style={at={(1,1)}, anchor=north east},
            bar shift=0cm
        ]
            \addplot+[fill=blue!40] coordinates {
                (SocNews,14247)
                (Entr,13684)
                (Gaming,8326)
                (PsychLife,4726)
                (HealthFit,3531)
                (Educ,2961)
                (Food,2805)
                (CultArts,2529)
                (Bus,2368)
                (TechSci,1538)
                (ParentFam,1064)
                (BeautFash,589)
                (MovBook,249)
            };
            \addplot+[fill=red!40] coordinates {
                (SocNews,4157)
                (Entr,4015)
                (Gaming,4133)
                (PsychLife,1101)
                (HealthFit,587)
                (Educ,590)
                (Food,370)
                (CultArts,584)
                (Bus,578)
                (TechSci,446)
                (ParentFam,266)
                (BeautFash,135)
                (MovBook,56)
            };
            \legend{Total Clips, Toxic Clips}
        \end{axis}
    \end{tikzpicture}
    \vspace{-4pt}
    \caption{Total and Toxic Clip Counts by Category.}
    \vspace{-10pt}
    \label{fig:clip_counts}
\end{figure}
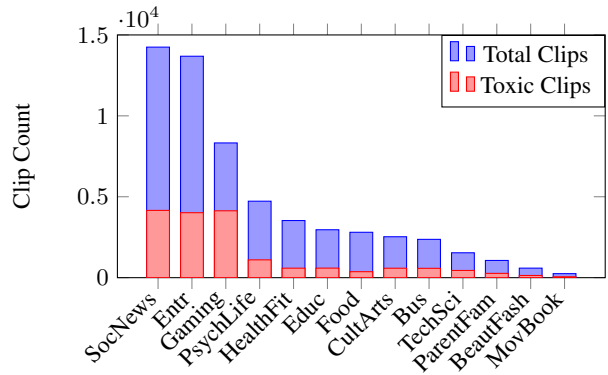

%% file: plots/toxicity_source_classification.tex
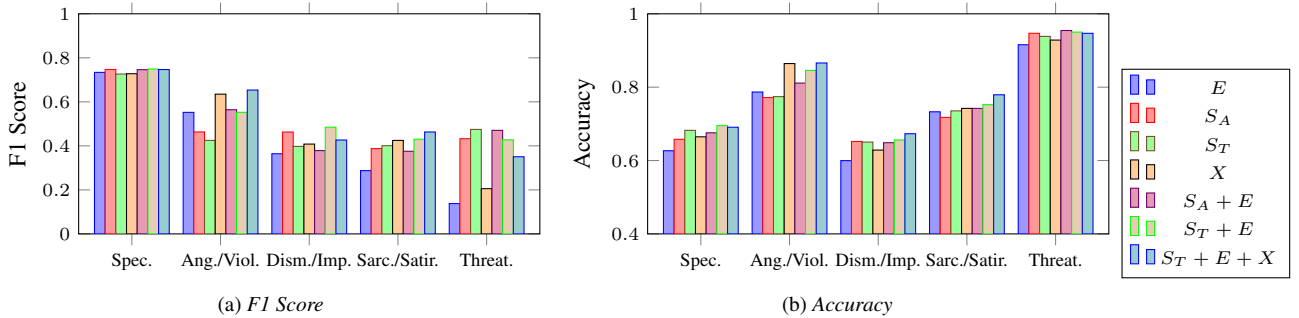
\begin{figure*}[!htbp]
\centering
\hspace{-40pt}
\begin{subfigure}[t]{0.45\textwidth}
    \centering
    \begin{tikzpicture}
    \begin{axis}[
        ybar,
        bar width=4pt,
        width=\linewidth,
        height=4.5cm,
        enlarge x limits=0.15,
        symbolic x coords={Spec. , Ang./Viol., Dism./Imp., Sarc./Satir., Threat.},
        xtick=data,
        ylabel={F1 Score},
        ylabel near ticks,
        ymin=0, ymax=1,
        tick label style={font=\scriptsize},
        label style={font=\small},
        legend to name=commonLegend, 
        legend style={font=\scriptsize, legend columns=1, at={(0,0)}, anchor=west}
    ]
        \addplot+[fill=blue!40, bar shift=-12pt] 
            coordinates {(Spec. ,0.7339) (Ang./Viol.,0.5522) (Dism./Imp.,0.3641) (Sarc./Satir.,0.2876) (Threat.,0.1379)};
        \addplot+[fill=red!40, bar shift=-8pt] 
            coordinates {(Spec. ,0.7472) (Ang./Viol.,0.4633) (Dism./Imp.,0.4629) (Sarc./Satir.,0.3875) (Threat.,0.4324)};
        \addplot+[fill=green!40, bar shift=-4pt] 
            coordinates {(Spec. ,0.7262) (Ang./Viol.,0.4248) (Dism./Imp.,0.3976) (Sarc./Satir.,0.4007) (Threat.,0.4748)};
        \addplot+[fill=orange!40, bar shift=0pt] 
            coordinates {(Spec. ,0.7278) (Ang./Viol.,0.6349) (Dism./Imp.,0.4081) (Sarc./Satir.,0.4248) (Threat.,0.2056)};
        \addplot+[fill=purple!40, bar shift=4pt] 
            coordinates {(Spec. ,0.7462) (Ang./Viol.,0.5642) (Dism./Imp.,0.3785) (Sarc./Satir.,0.3755) (Threat.,0.4705)};
        \addplot+[fill=brown!40, bar shift=8pt] 
            coordinates {(Spec. ,0.7493) (Ang./Viol.,0.5525) (Dism./Imp.,0.4848) (Sarc./Satir.,0.4302) (Threat.,0.4271)};
        \addplot+[fill=teal!40, bar shift=12pt] 
            coordinates {(Spec. ,0.7467) (Ang./Viol.,0.6536) (Dism./Imp.,0.4266) (Sarc./Satir.,0.4631) (Threat.,0.3505)};
        \legend{\(E\), \(S_A\), \(S_T\), \(X\), \(S_A+E\), \(S_T+E\), \(S_T+E+X\)}
    \end{axis}
    \end{tikzpicture}
    \caption{F1 Score}
    \label{fig:f1_toxic_source}
\end{subfigure}
\hspace{-10pt}
\begin{subfigure}[t]{0.45\textwidth}
    \centering
    \begin{tikzpicture}
    \begin{axis}[
        ybar,
        bar width=4pt,
        width=\linewidth,
        height=4.5cm,
        enlarge x limits=0.15,
        symbolic x coords={Spec. , Ang./Viol., Dism./Imp., Sarc./Satir., Threat.},
        xtick=data,
        ylabel={Accuracy},
        ylabel near ticks,
        ymin=0.4, ymax=1,
        tick label style={font=\scriptsize},
        label style={font=\small}
    ]
        \addplot+[fill=blue!40, bar shift=-12pt] 
            coordinates {(Spec. ,0.6267) (Ang./Viol.,0.7868) (Dism./Imp.,0.5998) (Sarc./Satir.,0.7329) (Threat.,0.9157)};
        \addplot+[fill=red!40, bar shift=-8pt] 
            coordinates {(Spec. ,0.6579) (Ang./Viol.,0.7716) (Dism./Imp.,0.6520) (Sarc./Satir.,0.7177) (Threat.,0.9469)};
        \addplot+[fill=green!40, bar shift=-4pt] 
            coordinates {(Spec. ,0.6823) (Ang./Viol.,0.7742) (Dism./Imp.,0.6503) (Sarc./Satir.,0.7354) (Threat.,0.9385)};
        \addplot+[fill=orange!40, bar shift=0pt] 
            coordinates {(Spec. ,0.6647) (Ang./Viol.,0.8644) (Dism./Imp.,0.6285) (Sarc./Satir.,0.7422) (Threat.,0.9284)};
        \addplot+[fill=purple!40, bar shift=4pt] 
            coordinates {(Spec. ,0.6756) (Ang./Viol.,0.8112) (Dism./Imp.,0.6486) (Sarc./Satir.,0.7422) (Threat.,0.9545)};
        \addplot+[fill=brown!40, bar shift=8pt] 
            coordinates {(Spec. ,0.6950) (Ang./Viol.,0.8458) (Dism./Imp.,0.6562) (Sarc./Satir.,0.7523) (Threat.,0.9502)};
        \addplot+[fill=teal!40, bar shift=12pt] 
            coordinates {(Spec. ,0.6908) (Ang./Viol.,0.8660) (Dism./Imp.,0.6731) (Sarc./Satir.,0.7793) (Threat.,0.9469)};
    \end{axis}
    \end{tikzpicture}
    \caption{Accuracy}
    \label{fig:acc_toxic_source}
\end{subfigure}
\hspace{-7pt}
\begin{subfigure}[t]{0.07\textwidth}
    \centering
    \pgfplotslegendfromname{commonLegend}
\end{subfigure}
\vspace{-8pt}
\caption{Performance of toxicity source classification models across different toxic sources. Plot (a) shows F1 scores and plot (b) shows Accuracy. The right panel shows the common legend of different embeddings used.}
\vspace{-10pt}
\label{fig:toxic_source_performance}
\end{figure*}

%% file: Tables/toxicity_classification.tex
\begin{table}[ht!]
    \centering
    \caption{Toxicity detection performance with existing models and different embedding configurations. All results are in \%. In each metric, the best performance is marked in \textbf{bold} and the second best is \underline{underlined}.}
    \vspace{-6pt}
    \label{tab:embedding_results}
    \begin{tabular}{lcccc}
        \toprule
        \textbf{Model} & \textbf{F1} & \textbf{Acc.} & \textbf{Prec.} & \textbf{Rec.} \\
        \midrule
        Mutox                & 29.41 & 67.19 & 45.79 & 21.66 \\
        Etox                 & 50.54 & 72.40 & 58.85 & 44.28 \\
        COLD\textsubscript{ETECTOR}  & 42.47 & 65.32 & 45.01 & 40.20 \\
        \midrule
        $E$                  & 47.96 & 72.18 & 57.49 & 41.14 \\
        $X$                  & 60.89 & 74.98 & 59.35 & 62.52 \\
        $S_A$                & 54.04 & 73.36 & 58.43 & 50.26 \\
        $S_T$                & 57.49 & 72.76 & 55.95 & 59.11 \\
        $S_A+E$              & 56.68 & 74.72 & 60.81 & 53.08 \\
        $S_T+E$              & \underline{62.47} & 73.76 & 56.35 & \textbf{70.09} \\
        $X+E$                & 60.62 & 74.02 & 58.47 & \underline{64.15} \\
        $X+S_A+E$            & 61.93 & \underline{76.52} & \underline{62.58} & 61.30 \\
        $X+S_T+E$            & \textbf{64.16} & \textbf{77.90} & \textbf{64.85} & 63.48 \\
        \bottomrule
    \end{tabular}
\end{table}